# Mapping and Monitoring Pollution Levels of Carbon Monoxide (CO) using Arduino and Location-Based Service

Purwantoro[1], Arip Solehudin[2], Nono Heryana[3]

[1,2] (Program Studi Teknik Informatika, FakultasIlmuKomputer, UnivesitasSingaperbangsaKarawang, Indonesia
Email: purwantoro.masbro@staff.unsika.ac.id, arip.solehudin@staff.unsika.ac.id)

[3] (Program StudiSistemInformasi, FakultasIlmuKomputer, Univesitas of SingaperbangsaKarawang, Indonesia
Email: nono@unsika.ac.id)

**Abstract:**

Air pollution is a serious issue in the world. Around 98% of cities with a population of over 100,000 people in low and middle-income countries do not meet air quality standards, while in high-income countries, the number has decreased by 52%. Industries and motor vehicles are the biggest contributors to carbon monoxide (CO). Air pollution is a severe issue in the world. The object of research to investigate is the detection of carbon monoxide (CO) levels in an area mapped based on the current CO levels along with the location coordinates via GPS and LBS technology with a microcontroller and sensor-based device. The results of testing the CO level detection devices found that high CO levels are in the afternoon with an average CO level of 49.59656 which means classified as dangerous if we are outdoors more than 30 minutes, can interfere with heart function.

## I. INTRODUCTION

Air pollution is a severe issue in the world. About 98% of cities with a population of over 100,000 people in low and middle-income countries do not meet air quality standards, whereas, in high-income countries, the number has decreased by 52% [1]. The presence of industrial activities and motor vehicles is the most significant contributor to air pollution. Industries and motor vehicles emit harmful substances to human health and the environment itself. Such as lead (Pb), suspended particulate matter (SPM), nitrogen oxides (NOx), hydrocarbons (HC), carbon monoxide (CO), and photochemical oxides (Ox). Motor vehicles account for almost 100% lead, 13-44% SPM, 71-89% HC, 34-73% NOx, and almost all carbon monoxide into the air [2]. CO is an odorless, colorless, tasteless, and non-irritating gas. CO is known to affect the work of the heart, central nervous system, fetus, and can affect the respiratory tract, which can cause oxygen deficiency and lead to death. Planting trees is a practical step in reducing CO [3]. Research related to detecting CO levels includes: to find out CO levels in the garage, [4], the impact of the number of motor vehicles on CO [5]. Analysis of CO concentrations [6] showed significant results. CO detection devices in the market and research above use a stand-alone system.

The object of research to investigate is the detection of carbon monoxide (CO) levels in an area mapped based on the current CO levels along with the location coordinates via GPS and LBS technology with a microcontroller and sensor-based device. Then the results of the CO record levels and location points are sent via GSM modules to the Android-based system to be processed and the effects of processing used for recommendations to authorized officials (village officials) and residents for planting trees that are useful in reducing carbon monoxide air pollution in the area under study. Research Location (lab/studio/ field) The location of the research is in the lab and the ground (village office) conducted by researchers and research members and assisted by experts in this research field. Other agencies involved are village officials who handle the environment and public health, their contribution to providing data and information related to research and experience, or constraints they face in environmental management. With data and information from village officials and related parties, it can be easier to develop this research system.

Mapping and monitoring system of pollution of carbon monoxide (CO) levels in an Arduino-based area and location-based service that was developed by the waterfall combined method for the domain of the software system with Rapid Control prototype for the realm of sensor network-based tool development with results and coordinates of test locations can be sent through mobile technology (android) to interested parties in the area under study with the urgency of research in areas around industrial areas and big cities the level of air pollution levels are higher so that the reduction of air pollution is immediately carried out.

## II. THEORETICAL REVIEW

### a. Carbon Monoxide

Carbon Monoxide (CO) is a colorless, odorless, tasteless gas consisting of one carbon atom that covalently bonded to one oxygen atom [1]. Carbon monoxide produced from incomplete combustion of carbon compounds, often occurring in internal combustion engines, formed when there is a lack of oxygen in the combustion process. Although toxic, CO plays an essential role in modern technology, which is a precursor of many carbon compounds.

According to WHO (1999), the limit of CO exposure in humans is 80 ppm for 15 minutes, 48 ppm for 30 minutes, 24 ppm for 1 hour, and 8 ppm for 8 hours. The threshold value is 25 ppm (PER.13 / MEN / X / 2011). Less than 25 ppm is still reasonably safe.

### b. Arduino

Arduino is a microcontroller board based on ATmega328 [7], which has 14 input/output pins of which 6 (six) pins can use as PWM outputs, sixanalog inputs, 16 MHz crystal oscillators, USB connections, power jacks, ICSP heads, and reset buttons. Arduino can support a microcontroller.

### c. MQ7 Sensor

A sensor is an equipment used to convert a physical quantity into an electrical quantity so that it can analyze with specific electrical circuits. MQ 7 is a gas sensor that is used in equipment to detect carbon monoxide (CO) gas in everyday life, industry, or cars. The feature of this MQ7 gas sensor is that it has a high sensitivity to carbon monoxide (CO), is stable, and has a long life.

### d. GPS (Global Positioning System)

GPS is a device or system that can be used to inform users on the surface of the earth based on satellites. Data sent from satellites in the form of radio signals with digital data. GPS can help show directions and is available free. GPS can be used anywhere in 24 hours. The position of the GPS unit will be determined based on the coordinates of the degrees of latitude and longitude in the operating system.

### e. Location-Based Service (LBS)

LBS is generally used to describe the technology used to find the location of the device used. LBS [9] is a service that is used to determine the position of the user and then uses that information to provide services and applications that are personalized using the help of GPS satellites to obtain accurate data positions.

## III. METHOD

The system development method in this research uses a combination of the waterfall method and the rapid control prototype (RCP).As for the methodology used can be seen in the following figure 1.

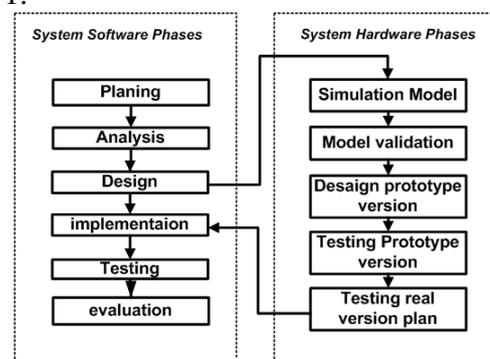

Figure 1. Research Stages

## IV. RESULTS & DISCUSSION

The results of this study are a system that can detect levels of carbon monoxide (CO) which is one of the toxic substances part of air pollution, the results of CO detection locations are mapped based on CO levels through the coordinates of the earth map location using GPS and made recommendations for planting trees to the village investigated in order to reduce pollution levels in the high carbon monoxide area. The following is an illustration of the results achieved in figure 2.

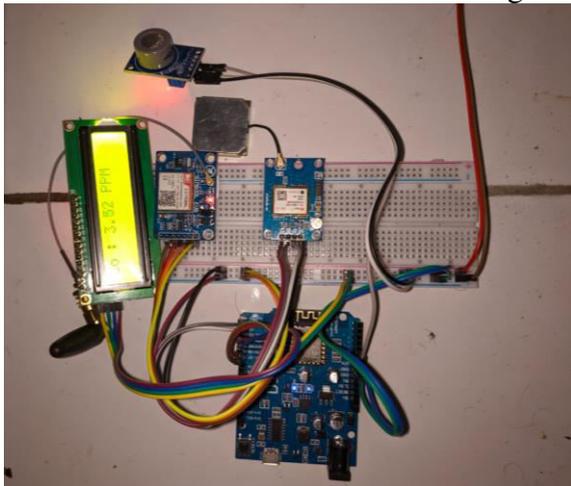

Figure 2. Component Assembly Results

As for the results of the implementation of the carbondioxide detector can be seen in the following figure 3.

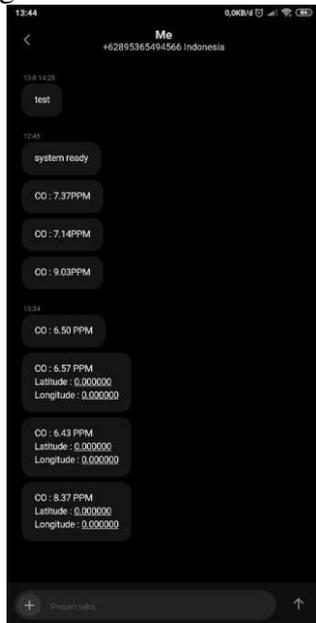

Figure 3. SMS Respond

The results of CO level recapitulation per location for five days when checking the value of CO levels after averaged.

a. CampusUnsika, Coordinates Coordinates: Lat-6.D6323799, Lng107.306427

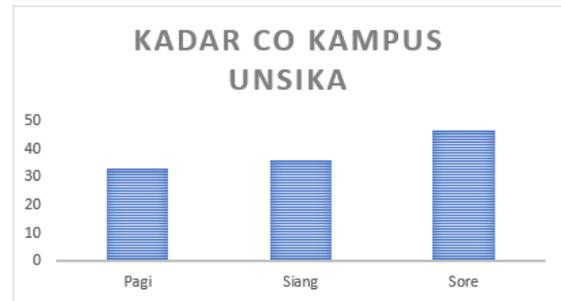

Figure 4. CO level in campus Unsika

| Time | CO levels | Description |
|---|---|---|
| Morning | 32,916 | Classified as dangerous if we are outside the room for more than 30 minutes |
| Noon | 36.0164 | Classified as dangerous if we are outside the room for more than 30 minutes |
| Afternoon | 46.7436 | Classified as dangerous if we are outside the room for more than 30 minutes, can interfere with the function of the heart |

b. CampusUBP, coordinates: Lat-6.323513, LNG 107.301137

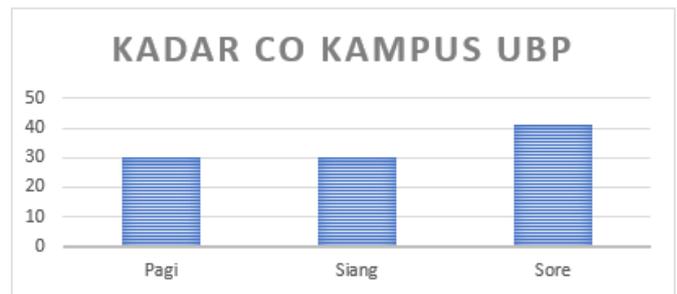

Figure 5. CO level in Campus UBP

| Time | CO levels | Description |
|---|---|---|
| Morning | 30,134 | Classified as dangerous if we are outside the room for more than 30 minutes |
| Noon | 30.3468 | Classified as dangerous if we |
| | | are outside the room for more than 30 minutes |
| Afternoon | 41.23 | Classified as dangerous if we are outside the room for more than 30 minutes, can interfere with the function of the heart |

c. Sky Bridge 2Galuh Mas, coordinates: Lat-6,327190, Lng107,291857

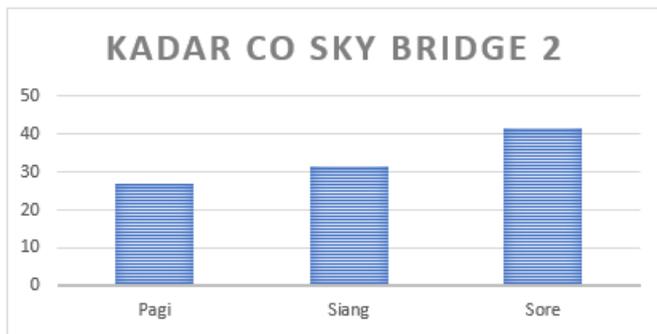

Figure 6. CO level in Sky Bridge 2 Galuh Mas

| Time | CO levels | Description |
|---|---|---|
| Morning | 26.9292 | A little dangerous if we are in the room for more than 45 minutes |
| Noon | 31.6068 | Classified as dangerous if we are outside the room for more than 30 minutes |
| Afternoon | 41.5176 | Classified as dangerous if we are outside the room for more than 30 minutes, can interfere with the function of the heart |

d. Traffic Light Bintang Alam, coordinates: Lat-6.332336, LNG 107.312257

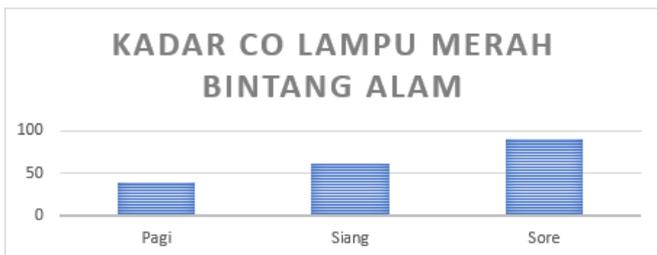

Figure 7. CO level in Traffic Light Bintang Alam

| Time | CO levels | Description |
|---|---|---|
| Morning | 39.0796 | Classified as dangerous if we are outside the room for more than 30 minutes |
| Noon | 62.3468 | Classified as dangerous if we are outside the room for more than 30 minutes, can make headaches |
| Afternoon | 89.79 | Very dangerous if inhalation of more than 15 minutes, will make it difficult to breathe |

e. MCD Galuh Mas, coordinates: Lat-6,329349, Lng107,296362

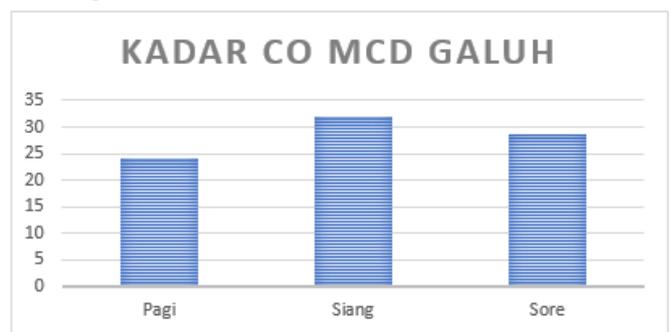

Figure 8. CO level in MCD Galuh Mas

| Time | CO levels | Description |
|---|---|---|
| Morning | 24,038 | Classified as safe for human |
| Noon | 32.0756 | Classified as dangerous if we are outside the room for more than 30 minutes |
| Afternoon | 28.7016 | A little dangerous if we are in the room for more than 45 minutes |

## V. CONCLUSIONS

Based on the results of testing, the CO level detection devices found that high CO levels are in the afternoon with an average CO level of 49.59656, which means classified as dangerous if we are outdoors more than 30 minutes, can interfere with heart function.

## ACKNOWLEDGMENT

This research was funded by the ministry of research and higher education through the PenelitianDosenPemula (PDP) Scheme.